# Graphical aids for relativistic optics

Bernhard Rothenstein[1], Corina Nafornita[2]

[1]"Politehnica" University of Timisoara, Dept. of Physics, Piata Regina Maria 1, Timisoara, Romania, e-mail: bernhard_rothenstein@yahoo.com.

[2]"Politehnica" University of Timisoara, Dept. of Communications, Bd. Parvan 2, Timisoara, Romania.

The paper presents a relativistic space-time diagram, which displays in true values the space (Cartesian and polar) and the time coordinates of the same event detected from two inertial reference frames in relative motion related by the Lorentz-Einstein transformations, the aberration angles and the Doppler shifted periods and wavelengths. We use it in order to illustrate the reflection of light on moving mirrors (horizontal and vertical) and the way in which a single observer could measure the length of a moving rod. It displays in true values the space-time coordinates of the same event generated by a light signal.

## 1. Introduction

Rosser [1] presents relativity optics at an introductory level. The topics involves the aberration of light effect, the transformation of plane waves in vacuum, the Doppler Effect, the reflection of light on a moving mirror and the visual appearance of rapidly moving objects. Relativists have developed relativistic diagrams usefully in relativistic kinematics, which display in true or in distorted values the space-time coordinates of the same event, as detected from two inertial reference frames in relative motion [2]. Constructing the diagram, the Authors use the Lorentz-Einstein transformations and transform a kinematics problem into a movie.

The purpose of our paper is to construct a relativistic diagram that visualizes the optical effects mentioned above, displaying in true values the magnitudes of the physical quantities introduced in order to characterize the effects we study as detected from two inertial reference frames in relative motion. We have in mind position vectors, time coordinates, wavelengths and periods, incidence and reflection angles, visual and radar detected appearances of rapidly moving objects.

We say that observers from two inertial reference frames in relative motion detect the same event if both of them agree that it takes place at the same point in space at times displayed by two clocks of the two reference frames instantly located at that point. The synchronization of the two clocks took place in theirs own rest frames in accordance with a procedure proposed by Einstein. If in a two space dimensions approach, the space-time coordinates of the same event are in one of the involved reference frames $E(x; y; t)$ whereas in the other they are $E'(x'; y'; t')$ then the Lorentz-Einstein transformations relate them.

Consider that in one of the involved inertial reference frames the corresponding observers perform an experiment. It is characterized by different events taking place at different points in space at different times. We say that observers from another inertial reference frame perform the same experiment if it involves the same events. In accordance with the relativistic postulate, the same experiments performed in the two reference frames should lead to the same results, which do not allow detecting if the reference frames are in a state of rest or in a state of uniform motion. The space-time coordinates of the same events are related by the Lorentz-Einstein transformations. Such an approach avoids paradoxes.

## 2. Constructing the relativistic diagram and finding out its abilities

The simplest way to construct the diagram is to start with a spherical mirror (circular in a two-space dimensions approach) with a point-like source located at its center. The center of the mirror coincides with the origin O' of its rest frame K'(X';O';Y') where we find the source of light S' as shown in Figure 1. We characterize an event by its space-time coordinates $E'(x'; y'; t') = E'(r' \cos\theta'; r' \sin\theta')$ where (x';y') and (r';θ') are the Cartesian and the polar coordinates of the point where the event takes place and t' the time when it takes place. Consider that S' emits light signals in all directions in space (event $O'_e(0;0;0)$). One of the light signals emitted along a direction, which makes an angle θ' with the positive direction of the O'X' axis (that is the way in which we define angles), generates arriving at the mirror the event $M'\left(x'; y'; t' = \dfrac{r'}{c}\right) = M'\left(r' \cos\theta'; r' \sin\theta'; t' = \dfrac{r'}{c}\right)$. The signal reflects itself and generates, returning to O', the event $O'_r\left(0; 0; \dfrac{2r'}{c}\right)$. Consider now the reference frame K(XOY). The axes of the two frames are parallel to each other, the OX(O'X') axes are overlapped and K' moves with constant velocity v in the positive direction of the overlapped axes. The clocks of K and K' read t=t'=0 when the origins O and O' are instantly located at the same point in space. In accordance with the relativistic postulate, the same events as detected from K are $O_e(0,0,0)$;

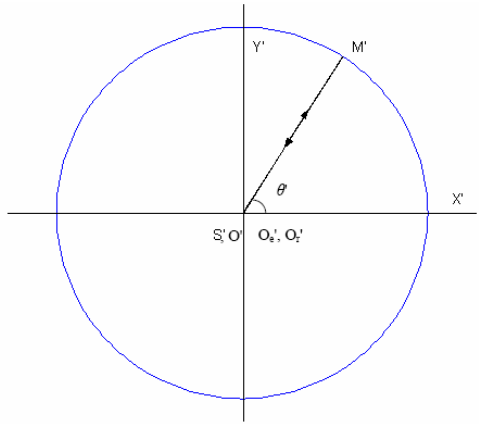

Figure 1. The circular mirror. Events $O'_e$, M' and $O'_r$ are associated with the emission of a light signal by a point-like source S' located at its center O', with its incidence on the mirror and with its return to O' respectively.

$$M\left(x; y; t=\frac{r}{c}\right) = M\left(r\cos\theta; r\sin\theta; t=\frac{r}{c}\right) \quad \text{and} \quad O_r\left(0; 0; \frac{2r}{c}\right). \text{ The Lorentz-Einstein}$$
transformations relate the space-time coordinates of the two events as
$$O_e'(0;0;0) \tag{1}$$

$$M'\left(x' = \gamma r\left(\cos\theta - \frac{v}{c}\right); y' = r\sin\theta; t' = \gamma\frac{r}{c}\left(1 - \frac{v}{c}\cos\theta\right)\right) \tag{2}$$

$$O_r'\left(x' = 2\gamma v\frac{r}{c}; y' = 0; t' = 2\gamma\frac{r}{c}\right) \tag{3}$$

if we use Cartesian coordinates and as

$$r = \frac{\gamma^{-1} r'}{1 - \frac{v}{c}\cos\theta} \tag{4}$$

$$\cos\theta = \frac{x}{r} = \frac{\cos\theta' + \frac{v}{c}}{1 + \frac{v}{c}\cos\theta'} \tag{5}$$

if we use polar coordinates. Equation

$$t = \frac{r}{c} = \frac{\gamma^{-1} t'}{1 - \frac{v}{c}\cos\theta} \tag{6}$$

relates the time coordinates, of events M and M'.
Taking into account that events M and M' are generated by light signals we can express r and r' as multiples of wave lengths ($\lambda, \lambda'$) and the time intervals (t-0) and (t'-0) as multiples of periods (T,T'), we have
$$r = n\lambda \tag{7}$$
$$r' = n\lambda' \tag{8}$$
and
$$t = nT \tag{9}$$
$$t' = nT' \tag{10}$$
where we have taken into account the invariance of the counted number of stable objects. The result is that wavelengths and periods transform as
$$\lambda = \lambda'\frac{\gamma^{-1}}{1 - \frac{v}{c}\cos\theta} \tag{11}$$

$$T = T'\frac{\gamma^{-1}}{1 - \frac{v}{c}\cos\theta}. \tag{12}$$

Equations (11) and (12) describe the optical Doppler effect [3]. We underline that T and T' are proper time intervals.

Multiplying both sides of (5) by c and taking into account that $c_x = c\cos\theta$; $c_y = c\sin\theta$ and $c'_x = c\cos\theta'$; $c'_y = c\sin\theta'$ represent the components of the velocity of the light signal in K and in K' respectively we obtain that equations

$$c_x = \frac{c'_x + v}{1 + \frac{v}{c^2}c'_x} \tag{13}$$

$$c_y = \frac{\gamma^{-1}c'_y}{1 + \frac{v}{c^2}c'_x} \tag{14}$$

relate them.
Equation

$$r' = R_0 \tag{15}$$

describes the circular mirror in its rest frame K'.

We obtain the relativistic diagram overlapping the circle r'=$R_0$ (15) with the ellipse (4) as we show in Figure 2 for the case when β = vc$^{-1}$ = 0.6.

The axes of the ellipse are $a = \gamma R_0$ and $b = R_0$ respectively. The distance between the two foci $F_1$ and $F_2$ of the ellipse is

$$\overline{F_1 F_2} = \frac{2R_0 \frac{v}{c}}{\sqrt{1 - \frac{v^2}{c^2}}} = 2e. \tag{16}$$

Equation [7]

$$r = \frac{p}{1 - \varepsilon \cos\theta} \tag{17}$$

describes an ellipse in polar coordinates. For our ellipse (4) with r'=$R_0$ we have

$$\varepsilon = \frac{e}{a} = \frac{v}{c} \tag{18}$$

$$p = \frac{b^2}{a} = R_0 \gamma^{-1} \tag{19}$$

resulting that

$$r = \frac{R_0 \sqrt{1 - \frac{v^2}{c^2}}}{1 - \frac{v}{c}\cos\theta} \tag{20}$$

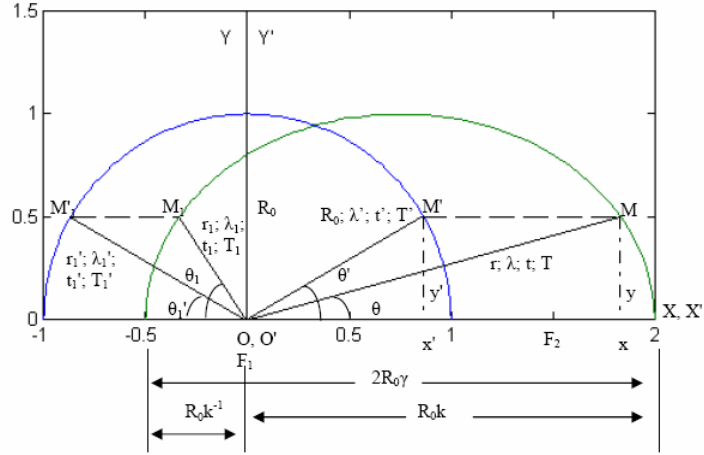

Figure 2. The relativistic space-time diagram for outgoing rays. We obtain it by overlapping the circular mirror with its apparent shape, which is an ellipse. Event M'($M_1^{\cdot}$) is associated with the incidence of the light ray in the mirror's rest frame whereas M($M_1$) located on the ellipse represents the same event detected from the stationary frame. Establishing the relationship between them, we have considered the invariance of distances measured perpendicular to the direction of relative motion. The diagram displays in true values the space (Cartesian and polar) and the time coordinates of events related by the Lorentz-Einstein transformations, the corresponding components of the velocity of light signals (outgoing) and the Doppler shifted wavelengths and periods. Constructing the diagram we have taken into account that the transformation equation for the length of the position vector and for the time coordinate have the same algebraic structure.

describes a genuine ellipse with all its well-known geometrical properties.

If event $M^{\cdot}(R_0, \theta^{\cdot})$ is located in our relativistic diagram on the circle, the same event detected from K is located on the ellipse. We have obtained the correspondence between the two events taking into account the invariance of distances measured perpendicular to the direction of relative motion. Constructing the diagram we have used the same units of length, in K and in K' as well.

The relativistic diagram we have just constructed displays in true values the corresponding polar angles θ and θ'. At corresponding scales, the segment $\overline{O^{\cdot}M^{\cdot}}$ could represent the length of the position vector r' of event M', its time coordinate t', the wave length λ' and the period T' of the radiation emitted by the source S' all measured in K'. The physical properties of the relativistic diagram ensure the fact that $\overline{OM}$ represents the corresponding physical quantities as measured in K. The projections of $\overline{O^{\cdot}M^{\cdot}}$ and $\overline{OM}$ on the axes O'X' and O'Y' and on OX and OY respectively could represent the components of the velocity of light signal or the Cartesian coordinates as well. As we see, the circular mirror in K' becomes when detected from K an elliptical mirror. The geometrical properties of the ellipse make that at point M where the incidence takes place when detected from K, the ray, which comes from the focal point $F_1$, reflects itself towards the other focal point $F_2$. The normal to the ellipse at point M is the bisecting line of the angle made by the incident ray with the reflected one, in accordance with the reflection law (Figure 3).

At a time $t^{\cdot} = r^{\cdot}/c$, all the points of the spherical mirror become luminous and the reflected rays return to the center of the mirror at a time $t^{\cdot} = 2r^{\cdot}/c$. We make now, without loosing in generality, a shift in time considering that the reflection on the mirror takes place at a time and that the reflected rays return to O' at a zero time. An event, which takes place on the mirror has the space-time coordinates $M'\left(r'\cos\theta'; r'\sin\theta'; -\dfrac{r'}{c}\right)$ in K' and $M\left(r\cos\theta; r\sin\theta; -\dfrac{r}{c}\right)$ in K.

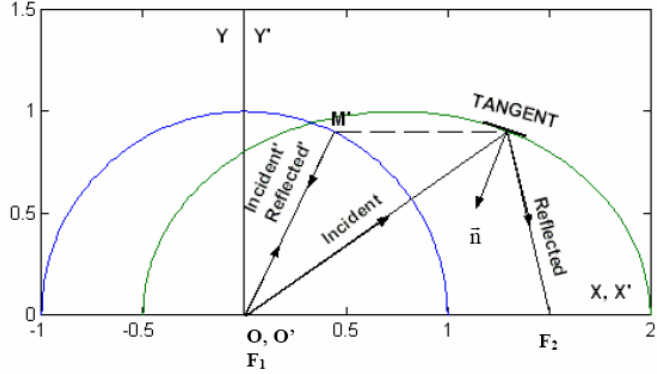

Figure 3. In the rest frame of the mirror the incident ray starts from the center of the mirror O', reflects itself on the mirror and returns to its starting point. Detected from the stationary frame the circular mirror becomes an elliptic one. The ray starts from its focal point $F_1$ and returns to the other focal point $F_2$. The geometric properties of the ellipse make that at the reflection point the normal is the bisecting line of the angle made by the incident and the reflected rays, in accordance with the reflection law convincing us that it holds in all inertial reference frames in relative motion.

The event associated with the return of the reflected ray at O' has the space-time coordinates $O'(0;0;0)$ in K' and $O(0;0;0)$ in K. In accordance with the Lorentz-Einstein transformations we have

$$x' = \gamma r(1 + \frac{V}{c}\cos\theta) \tag{21}$$

$$y' = r\sin\theta \tag{22}$$

$$r' = \gamma r(1 + \frac{V}{c}\cos\theta) \tag{23}$$

$$t' = -\gamma\frac{r}{c}(1 + \frac{V}{c}\cos\theta) \tag{24}$$

$$\cos\theta' = \frac{\cos\theta + \dfrac{V}{c}}{1 + \dfrac{V}{c}\cos\theta} \tag{25}$$

and

$$r = \gamma^{-1}\frac{r'}{1 + \dfrac{V}{c}\cos\theta}. \tag{26}$$

If Equation (4) works in the case of *"outgoing" (incident)* rays, Equation (26) holds in the case of *"incoming" (reflected)* rays.
Equations

$$\lambda = \gamma^{-1} \frac{\lambda'}{1+\frac{v}{c}\cos\theta'} \qquad (27)$$

and

$$T = \gamma^{-1} \frac{T'}{1+\frac{v}{c}\cos\theta'} \qquad (28)$$

describe the Doppler effect in the case of incoming rays. We obtain the corresponding relativistic diagram by overlapping the circle r'=R$_0$ with the ellipse (26) as we show in Figure 4. The diagram displays in true values the corresponding values of the physical quantities involved in the studied experiment as detected from K' and K respectively (r,x,y,θ,λ,T) and (r',x',y',θ',λ',T').

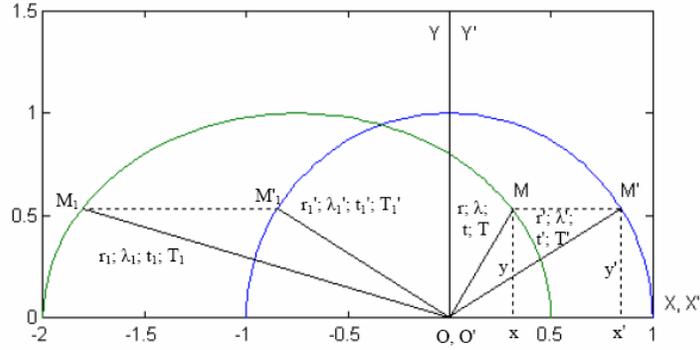

Figure 4. The relativistic space-time diagram for incoming ray. Event M'($M'_1$) is associated with the fact that the reflected ray starts from a point on the circular mirror being received at O' at a time t'=0. The corresponding event is M(M$_1$). The diagram displays in true values the space (Cartesian and polar) and time coordinates of events related by the Lorentz-Einstein transformations, the components of the velocity of light signals, the Doppler shifted wavelengths and periods, as detected from the two involved reference frames in relative motion.

In the case of the outgoing rays as in the case of the incoming ones, we do not detect relativistic effects along a well-defined direction. Imposing the condition r = r' we obtain in the case of the outgoing rays for the angle θ$_0$ along which we do not detect relativistic effects

$$\cos\theta_0 = \frac{1-\gamma^{-1}}{v/c} \qquad (29)$$

whereas in the case of the incoming rays we obtain

$$\cos\theta_0 = \frac{\gamma^{-1}-1}{v/c}. \qquad (30)$$

The relativistic diagram we have constructed so far displays those directions as the line which joins the overlapped origins O and O' with the point where the circle and ellipse intersect each other as we show in Figures 5a and 5b.

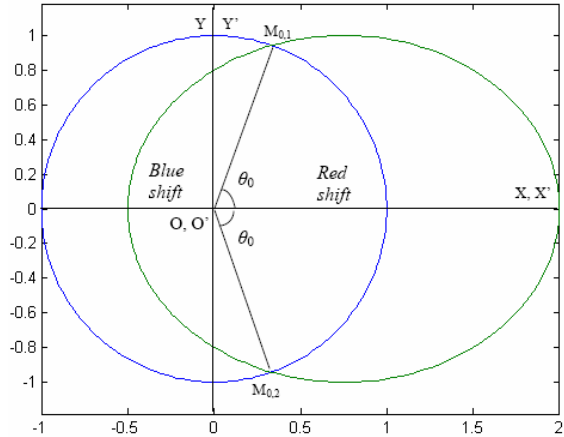

Figure 5a. The intersection of the circle with the ellipse determines the events $M_{0,1}(R,\theta_0)$ and $M_{0,2}(R,-\theta_0)$. Along the directions $OM_{0,1}$ and $OM_{0,2}$ we do not detect relativistic effects for the outgoing rays. The mentioned directions separate the directions along which a red shift takes place ($\lambda>\lambda'$) from the directions along which a blue shift takes place ($\lambda<\lambda'$).

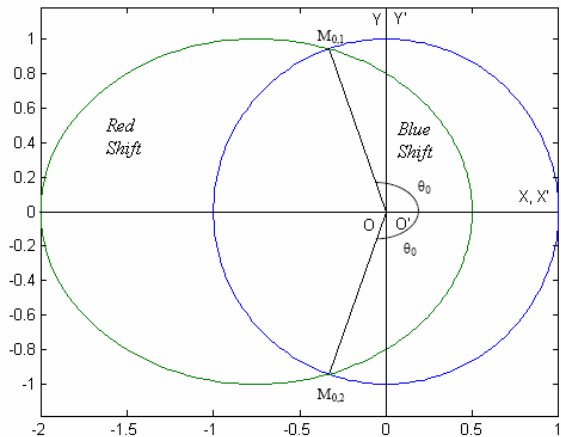

Figure 5b. The intersection of the circle with the ellipse determines the events $R_{0,1}(R,\theta_0)$ and $R_{0,2}(R,-\theta_0)$. Along the directions $OM_{0,1}$ and $OM_{0,2}$ we do not detect relativistic effects for incoming rays. The mentioned directions separate the directions along which a red shift takes place ($\lambda>\lambda'$) from the directions along which a blue shift takes place ($\lambda<\lambda'$).

## 3. Reflection of light on moving mirrors

We know that the reflection of light in the rest frame of the mirror takes place in accordance with the reflection law:
.Incident and reflected rays are located in the same plane,
..Incidence and reflection angles are equal to each other,
The reflection of light on a mirror involves an incoming ray (incident) and an outgoing ray (reflected). That peculiarity suggests that we could solve the problem of reflection on moving mirrors, overlapping the two relativistic diagrams constructed above in a single one.

## 3.1 Horizontal mirror located on the overlapped axes, point like source or incident plane wave

The plane mirror under consideration is on the overlapped axes OX(O'X') at rest in K' (Figure 6). We present an incident ray (incident)' and the corresponding reflected ray (reflected)' assuming that the reflection law holds in that frame. The rules of handling the relativistic diagram enable us to find out the corresponding incident (incident) and reflected rays (reflected) as detected from the K reference frame. If $\alpha_0$ and $(\pi-\alpha_0)$ represent the angles made by the reflected and the incident rays respectively with the overlapped axes as detected from the rest frame of the mirror (K'), then when detected from K the reflected ray propagates in the direction

$$tg\alpha_r = \frac{\gamma^{-1}\sin\alpha_0}{\cos\alpha_0 + \frac{V}{c}} \qquad (31)$$

whereas the incident ray propagates in a direction

$$tg\alpha_i = -\frac{\gamma^{-1}\sin\alpha_0}{\cos\alpha_0 + \frac{V}{c}} \qquad (32)$$

The symmetry of the results ($\alpha_i=\pi-\alpha_r$) convinces us that *the reflection law holds in the K reference frame as well*. Taking into account the handling rules of the relativistic diagram, we see that it leads to the same result.

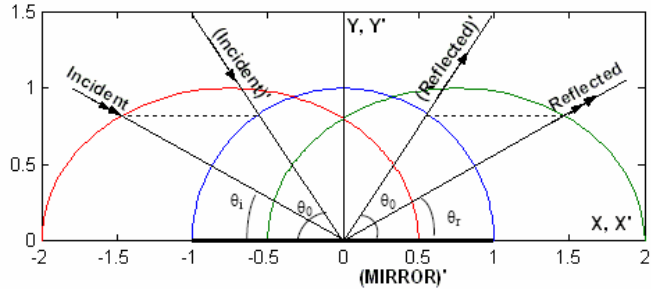

Figure 6. The considered mirror is on the overlapped axes OX(O'X') and at rest in the moving reference frame. The reflection law holds in the rest frame of the mirror and so the (incident') and the (reflected') rays make the same angle with the normal to the mirror. The relativistic space-time diagram tells us that when detected from the stationary frame, the reflection law works as well, the (incident) and (reflected) rays making the same angles with the normal to the mirror.

The (incident)' ray we have considered so far can be one of the rays in a plane wave in oblique incidence or one of the rays emitted by a point-like source S'. In the first case, the incidence conditions are the same at all the points of the horizontal mirror. In the second case, the incidence conditions are different at its different points but the reflection law holds in both cases in all inertial reference frames in relative motion.

We reverse now the paths of the rays we have considered so far (Figure 7). The rules of handling the relativistic diagram tell us that as detected from K the incident and the reflected rays change theirs previous paths. Even if the reflection law still works in K, the *reversibility* of the paths does not hold in the case of relativistic mirrors.

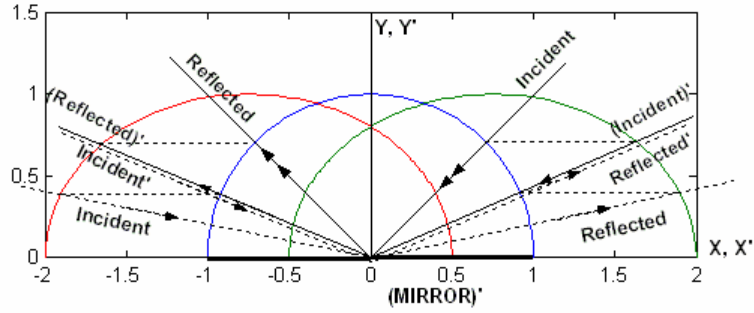

Figure 7. The space-time diagram illustrates the relativistic invariance of the paths of the incident and of the reflected rays in the rest frame of the mirror. If in that reference we reverse the direction of the rays then in the stationary reference frame the corresponding rays do not overlap. We can say that the invariance of the paths of the incident and of the reflected rays is not a relativistic invariance.

## 3.2 Vertical mirror and plane wave in oblique incidence

We consider one of the rays of a plane wave in oblique incidence on a vertical mirror located on the O'Y' axis of its rest frame K' at point O' at a time t=t'=0 (Figure 8). It reflects itself in accordance with the reflection law. Using the relativistic diagram, we find out easily the paths of the incident and of the reflected rays as detected from K. In accordance with the formulas, which describe the aberration of light effect we have for the incident ray

$$\operatorname{tg} \alpha_i = -\frac{\gamma^{-1} \sin \alpha_0}{\cos \alpha_0 - \frac{v}{c}} \qquad (33)$$

whereas for the reflected ray we have

$$\operatorname{tg} \alpha_r = \frac{\gamma^{-1} \sin \alpha_0}{\cos \alpha_0 + \frac{v}{c}} \qquad (34)$$

Combining Equations (33) and (34) we obtain

$$\frac{\operatorname{tg} \alpha_r}{\operatorname{tg} \alpha_i} = -\frac{\cos \alpha_0 - \frac{v}{c}}{\cos \alpha_0 + \frac{v}{c}} \qquad (35)$$

an equation which describes the reflection law on the moving mirror in two space dimensions. We derive for it more sophisticated equations.

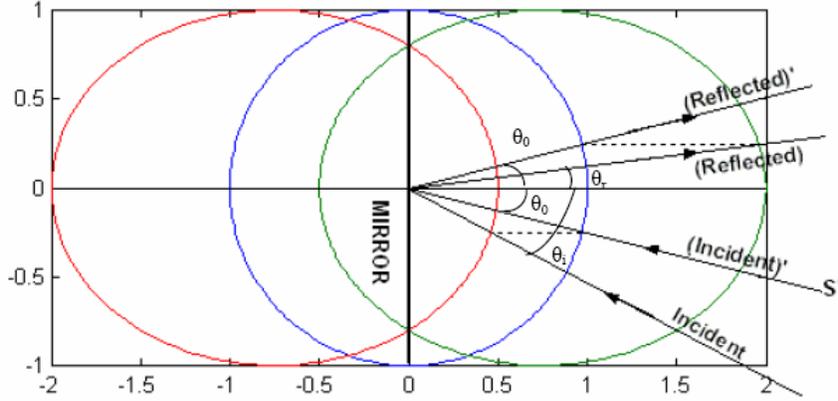

Figure 8. The reflection on a vertical moving mirror illustrated by the relativistic space-time diagram.

Starting with the obvious equations

$$\cos\alpha_r = \frac{\cos\alpha_0 + \frac{v}{c}}{1 + \frac{v}{c}\cos\alpha_0} \tag{36}$$

$$\cos\alpha_i = \frac{\cos\alpha_0 - \frac{v}{c}}{1 - \frac{v}{c}\cos\alpha_0} \tag{37}$$

we obtain eliminating $\alpha_0$ between those two equations [1]

$$\frac{\cos\alpha_r}{1 - \frac{v}{c}\cos\alpha_r} = \frac{\cos\alpha_i}{1 + \frac{v}{c}\cos\alpha_i} \tag{38}$$

We also have

$$\cos\alpha_0 = \frac{\cos\alpha_i - \frac{v}{c}}{1 - \frac{v}{c}\cos\alpha_i} \tag{39}$$

and

$$\cos\alpha_0 = \frac{\cos\alpha_r + \frac{v}{c}}{1 + \frac{v}{c}\cos\alpha_r} \tag{40}$$

from where with some algebra we obtain

$$\cos\alpha_r = \frac{\cos\alpha_i\left(1 + \frac{v^2}{c^2}\right) - 2\frac{v}{c}}{1 - 2\frac{v}{c}\cos\alpha_i + \frac{v^2}{c^2}} \tag{41}$$

The results obtained so far suggest that the reflection law does not hold in the K frame. In order to clarify the problem we consider a part of the vertical mirror of length 2a and two rays of the plane wave in oblique incidence arriving at the lower (1') and at the upper (2') of its ends (Figure 9).

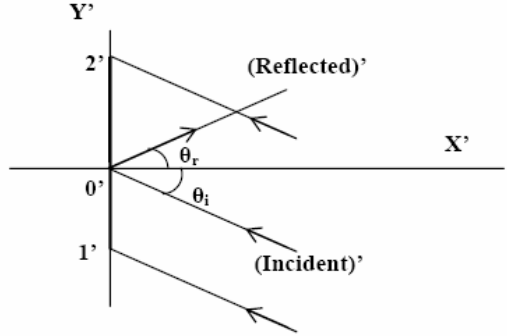

Figure 9. Calculating the apparent shape of the moving mirror.

They create arriving at the mirror the events $1'(0;-a;-\frac{a}{c}\sin\alpha_0)$; $O'(0;0;0)$; $2'(0;a;\frac{a}{c}\sin\alpha_0)$. The same events detected from K have the space coordinates $1(-\gamma\frac{v}{c}a\sin\alpha_0;-a)$; $O(0,0)$; $2(\gamma v\frac{a}{c}\sin\alpha_0;a)$. The result is that as detected from K the mirror appears rotated from its initial position with an angle

$$\tg\theta = \frac{y_2 - 0}{x_2 - 0} = \frac{\gamma^{-1}}{\frac{v}{c}} \tag{42}$$

Using the derived values of the angles $\alpha_i$, $\alpha_r$ and $\theta$ we can show that the reflection law works in the stationary reference frame as well.

## 4. Transforming monochromatic radiation in a polychromatic one via the Doppler effect.

Consider that the point-like source S' located at the origin O' of its rest frame K' emits for the observers of that frame a monochromatic radiation. The problem is to find out what does become that radiation for the observers of K. The relativistic diagram (Figure 5a and 5b) tells us that the wavelength of the radiation emitted inside the angle $2\alpha_0$ increases whereas the wavelength of the radiation emitted inside the angle $2(\pi - \alpha_0)$ decreases. We say that in the first case the radiation is *red shifted* and that in the second case it is *blue shifted*. As we have seen above, the angle defines the direction along which we do not detect relativistic effects. If the source S' emits N photons uniformly in all directions, then

$$N_r = \frac{N}{2\pi} 2\cos^{-1}\left(\frac{1-\gamma^{-1}}{\frac{v}{c}}\right) \tag{43}$$

represents the number of the red shifted photons, whereas

$$N_b = N - N_r = \frac{N}{\pi}\left(1 - \cos^{-1}\frac{1-\gamma^{-1}}{\frac{v}{c}}\right) \qquad (44)$$

represents the number of the blue shifted photons. The probability to detect a red shifted photon is higher the probability to detect a blue shifted one.